# Antiferromagnetic Skyrmion-Based Logic Gates Controlled by Electric Currents and Fields

Xue Liang,[1, 2, †] Jing Xia,[1, †] Xichao Zhang,[2] Motohiko Ezawa,[3] Oleg A. Tretiakov,[4] Xiaoxi Liu,[5] Lei Qiu,[6] Guoping Zhao,[1, 7 *] and Yan Zhou [2, *]

[1]*College of Physics and Electronic Engineering, Sichuan Normal University, Chengdu 610068, China*

[2]*School of Science and Engineering, The Chinese University of Hong Kong, Shenzhen, Guangdong 518172, China*

[3]*Department of Applied Physics, University of Tokyo, Hongo 7-3-1, Tokyo 113-8656, Japan*

[4]*School of Physics, The University of New South Wales, Sydney 2052, Australia*

[5]*Department of Electrical and Computer Engineering, Shinshu University, Wakasato 4-17-1, Nagano 380-8553, Japan*

[6]*The Center for Advanced Quantum Studies and Department of Physics, Beijing Normal University, Beijing 100875, China*

[7]*Center for Computational Sciences, Sichuan Normal University, Chengdu, 610068, China*

(Dated: July 28, 2021)

†These authors contributed equally.

Corresponding Authors
*(G.Z.) E-mail: zhaogp@uestc.edu.cn
*(Y.Z.) E-mail: zhouyan@cuhk.edu.cn

# ABSTRACT

Antiferromagnets are promising materials for future spintronic applications due to their unique properties including zero stray fields, robustness versus external magnetic fields and ultrafast dynamics, which have attracted extensive interest in recent years. In this work, we investigate the dynamics of isolated skyrmions in an antiferromagnetic nanotrack with a voltage-gated region. It is found that the skyrmion can be jointly controlled by the driving current and the voltage-controlled magnetic anisotropy gradient. We further propose a design of logic computing gates based on the manipulation of antiferromagnetic skyrmions, which is numerically realized combining several interactions and phenomena, including the spin Hall effect, voltage-controlled magnetic anisotropy effect, skyrmion-skyrmion interaction, and skyrmion-edge interaction. The proposed logic gates can perform the basic Boolean operations of the logic AND, OR, NOT, NAND and NOR gates. Our results may have a great impact on fundamental physics, and be useful for designing future non-volatile logic computing devices with ultra-low energy consumption and ultra-high storage density.

**KEWORDS:** *Magnetic skyrmion, antiferromagnet, logic gate, spintronics, micromagnetics*

Antiferromagnetic (AFM) materials are abundant in nature [1-6], which include certain oxides, sulfides and halogen compounds of elements of the iron group, such as FeO, MnO, $Cr_2O_3$, FeS, and $FeF_2$. In an AFM system, there are two interpenetrating sublattices, and the magnetic moments of both sublattices are equal in magnitude and coupled by the inter-sublattice exchange interaction [7]. An ideal AFM system is insensitive to external magnetic fields and has no stray field [4,5,8-10]. Such a feature may improve the stability and reliability of spintronic devices made of AFM materials, in particular against the detrimental effect of magnetic field perturbations [4]. However, the insensitivity of AFM materials to external magnetic fields also makes them hard to be controlled by magnetic fields, which is a method usually applied to ferromagnetic (FM) systems. Due to the zero net magnetization, it is also difficult to observe the AFM spin textures by some conventional experimental techniques, such as using the magneto-optical Kerr effect (MOKE) microscope [11].

For these reasons, although AFM materials were discovered early in the 1930s[1], they are not immediately favored by researchers until the observation or realization of the spin pumping effect [8,9], spin Hall effect [2,12], spin Seebeck effect [13,14], and spin-orbit torques [2,15,16] in AFM devices in recent years. Nowadays, it has also been found that in the spin-orbitronics [8], an emerging direction of spintronics, the magnetic order in AFM materials can be electrically controlled and read-out [2,10,17-19], which provide effective ways to realize AFM spintronic devices.

On the other hand, recent spintronics research has focused on AFM skyrmions [4,20-29], which are topologically nontrivial spin configurations and are promising for building future spintronic applications. In particular, there are many theoretical works devoted to studying the logic computing devices [22,30-45] based on various dynamic behaviors of skyrmions in geometrically confined racetracks, such as using the reversible conversion between the skyrmion and the domain wall [22,30], skyrmion Hall effect [31,36,37,39] and the skyrmion annihilation [38]. Recently, Z.R. Yan et al.[40] have also designed a skyrmion-based programmable logic device with complete Boolean logic functions. The AFM skyrmions can be regarded as a combination of two antiferromagnetically exchange-coupled FM skyrmions, where one sublattice skyrmion has a topological charge of +1 and the other has a topological charge of –1 [20,21]. Compared with FM skyrmions, AFM skyrmions hold three advantages: 1) the vanishing net magnetization and zero stray filed could enable AFM skyrmions to robustly resist the magnetic perturbations and hence make them potential components for satellite and aircraft electronics [5]; 2) AFM skyrmions can move along the direction of driving force without showing the skyrmion Hall effect [20-23,28]; 3) it has been theoretically predicted that AFM skyrmions could move at an ultra-high speed driven by spin currents [20-22,28].

In this work, we propose a design of Boolean logic computing gates based on AFM skyrmions, in which the presence and absence of AFM skyrmions represent the binary digit "1" and "0", respectively. The proposed logic gates can realize the basic Boolean logic operations of AND, OR, NOT, NAND and NOR gates by using the spin Hall effect [9], the spin transfer torque, the voltage-controlled magnetic anisotropy (VCMA) effect [46,47], the skyrmion-skyrmion interaction [48] and the skyrmion-edge interaction [49] simultaneously. It should be noted that the NAND gate is a universal gate and any other logic gates can be implemented by a combination of NAND gates in principle [35]. In our model, the spin-polarized current and the VCMA gradient are used as driving forces, which control the velocity and trajectory of AFM skyrmions in a joint manner.

Figure 1 schematically illustrates our proposed logic computing gates based on the manipulation of AFM skyrmions, including a cross-shaped AFM nanotrack that can be made by lithography technique [50] and three magnetic tunnel junctions (MTJs) as input and output heads. The heavy metal layer beneath the AFM layer is used to generate spin currents by using the spin Hall effect [51]. The insulator and electrode structures placed on the transverse AFM nanotrack is used to control the perpendicular magnetic anisotropy (PMA) within the transverse branch. As shown in Fig. 1(b), the wedge-shaped insulator will give rise to a linear change in PMA due to the VCMA effect [46,47], leading to a PMA gradient. Namely, once the gate voltage is applied, the PMA depends on the thickness of the insulator spacer and can be described by the linear equation [37,52] $K_v = K_0 + \xi V_g/td$, where $\xi$ is the VCMA coefficient fixed at 48 fJ·V$^{-1}$m$^{-1}$ in this work, $V_g$ is the applied voltage in the gated region, $t = 0.4$ nm and $d = 1.3$ nm are the thickness of the magnetic layer and the insulator layer, respectively. As reported in Ref. [23], the PMA gradient can drive AFM skyrmion into motion along a nanotrack effectively, similar to the spin-polarized current and temperature gradient [21,53]. It should be noted that the PMA of the transverse nanotrack is not entirely modified by the gate voltage, and only the region indicated by the blue dashed box in Fig. 1(c) is voltage gated. The PMA of the nanotrack edges remains to be unchanged during all operations, which could better confine skyrmions within the nanotrack and prevent the annihilation of skyrmions at edges, especially at the junction of the transverse and longitudinal branches.

Here, all micromagnetic simulations are performed by using the Object-Oriented MicroMagnetic Framework (OOMMF) with extended Dzyaloshinskii-Moriya interaction (DMI) [54,55] module [56]. The time-dependent magnetization dynamics is governed by the well-known Landau-Lifshitz-Gilbert (LLG) equation with spin transfer torque (STT):

$$\frac{d\mathbf{m}}{dt} = -|\gamma|\mathbf{m} \times \mathbf{H}_{eff} + \alpha \left(\mathbf{m} \times \frac{d\mathbf{m}}{dt}\right) + \boldsymbol{\tau}_{\text{STT}}, \qquad (1)$$

where $\mathbf{m}(\mathbf{r}) = \mathbf{M}(\mathbf{r})/M_s$ is the normalized magnetization, $\mathbf{H}_{eff} = -\delta\varepsilon/(\mu_0\delta\mathbf{m})$ is the effective field associated with various energies, such as the exchange energy, the magnetocrystalline anisotropic energy and the DMI energy, $\gamma$ and $\alpha$ are the gyromagnetic ratio and damping coefficient, respectively. Rewriting the STT limited to driving current leads to

$$\boldsymbol{\tau}_{\text{STT}} = \gamma H_D \mathbf{m} \times (\mathbf{m} \times \mathbf{m}_p), \qquad (2)$$

where $H_D = \hbar P j/(2\mu_0 e M_s t)$ is the effective field induced by the applied driving current, $P$ and $\mathbf{m}_p$ are, respectively, the spin polarization rate with default value of 0.4 and the unit electron polarization direction set as $\mathbf{e}_y$ to drive the AFM skyrmion, $\hbar$ is the Planck constant, $\mu_0$ is the permeability of vacuum, $e$ is the electron charge, $j$ represents the driving current density and $t$ is the thickness of magnetic film.

We consider a thin film with the easy-axis magnetic anisotropy perpendicular to the $xy$-plane and the DMI originating from the broken inversion symmetry at the interface. The interfacial DMI is introduced to stabilize the Néel-type skyrmion [23]. Therefore, the total AFM energy density $\varepsilon$ is a function of $\mathbf{m}$ specified by [20,22,27]

$$\varepsilon = A\left[(\nabla m_x)^2 + (\nabla m_y)^2 + (\nabla m_z)^2\right] - K_0 m_z^2 + D[m_z(\nabla \cdot m) - (m \cdot \nabla)m_z], \qquad (3)$$

where $A$ and $K_0$ are the exchange stiffness and the PMA constants, respectively, and the last term denotes the micromagnetic energy density associated with the DMI. More numerical simulation details can be found in the supplementary material.

Before studying the AFM-skyrmion-based logic gates, we investigate the basic dynamic behavior of a single AFM skyrmion in a straight nanotrack with a voltage-gated region, where the constant voltage is applied to modify the local PMA magnitude. As shown in Fig. 2(a-c), the black shaded box denotes the voltage-gated region. Here, we consider that the PMA value $K_v$ in the voltage-gated region is constant but smaller than that outside of the voltage-gated region. The PMA difference is defined as $\Delta K$, i.e., $\Delta K = K_0 - K_v$. As a consequence, the voltage-gated region serves as a potential well. Similar to FM skyrmions [57], the possibility of the AFM skyrmion passing through the voltage-gated region mainly depends on the magnitude of the driving current density as well as the anisotropy difference between the voltage-gated and non-voltage-gated regions.

As shown in Fig. 2(a-b), for a same value of $\Delta K$, the driving force provided by a small spin current is insufficient for the AFM skyrmion to overcome the energy barrier at the right boundary of the voltage-gated region, so that the skyrmion is trapped eventually. However, if a large current density is applied, the skyrmion will pass through the voltage-gated region. Besides, the diameter of the AFM skyrmion will increase in the voltage-gated region due to the smaller value of PMA. Therefore, when $\Delta K$ is larger than certain threshold, the skyrmion is not stable in the voltage-gated region and could be converted into a domain wall pair (see Fig. 2(c)). Such a conversion between a skyrmion and a domain wall pair can be used as a procedure for realizing the duplication of skyrmion [30,58-61]. The illustration of an AFM skyrmion is given in Fig. 2(d). Fig. 2(e) shows the working window of the current-driven motion of a single AFM skyrmion in the voltage-gated nanotrack. It can be seen that the critical driving current density $j_c$ required to drive the AFM skyrmion passing through the voltage-gated region is proportional to $\Delta K$. However, when $\Delta K$ is larger than $0.3 \times 10^5$ J m$^{-3}$ in our model, the skyrmion will be broken when it moves into the voltage-gated region.

We proceed to consider the case when two AFM skyrmions are moving in the nanotrack. Due to the presence of the skyrmion-skyrmion repulsion [48,49], we find that four different motion behaviors of skyrmions could occur. As shown in Fig. 3(a-c), similar to the case of only one skyrmion in the nanotrack, the AFM skyrmions will be destroyed if the PMA in the voltage-gated region is smaller than a certain threshold. However, when the two AFM skyrmions are stable in the voltage-gated region, they are intended to move away from each other due to the skyrmion-skyrmion repulsion. When the driving current density is small, both two AFM skyrmions cannot pass through the right boundary of the voltage-gated region due to the insufficient driving forces provided by the spin current. When the applied current density is slightly increased but still smaller than the critical current density $j_c$ (i.e., the critical current density for a single AFM skyrmion to pass through the voltage-gated region), the right skyrmion trapped at the right boundary will be pushed out of the voltage-gated region by the left skyrmion (see Fig. 3(b)). Indeed, the left skyrmion will stay in the voltage-gated region as long as the current density is smaller $j_c$. If the current density is larger than $j_c$, both two AFM skyrmions can pass through the voltage-gated region (see Fig. 3(c)).

Figure 3(d) shows the working window of the current-induced motion of two AFM skyrmions in a nanotrack. Compared with the case of a single AFM skyrmion (see Fig. 2(e)), when the driving current density is smaller than $j_c$, the original phase is divided into two phases. The phenomenon that only one AFM skyrmion is outputted when two AFM skyrmions are inputted could be applied in the transmission, modification and replacement of data in the skyrmion-based racetrack memory [62], as well as in the skyrmion-based logic computing gates, which we will discuss below. The relationship between the transition speed and the voltage applied has also been shown in the supplementary material.

In our proposed model of logic computing gates, the spin-polarized current is applied in the longitudinal nanotrack (i.e., along the *x* direction), while the transverse branch has the VCMA gradient without driving current. Hence, the AFM skyrmions on both nanotracks can be driven into motion, and the junction area of the longitudinal and transverse branches is a voltage-gated region with lower PMA. As shown in Fig. 4(a-b), assuming that only one AFM skyrmion is initially created in the input A side, it will first move to the junction area driven by the spin current. If the driving current density is smaller than $j_c$, the skyrmion will continue to move to the lower terminal of the transverse nanotrack due to the VCMA gradient, and thus no skyrmion can be detected at the output side (see Fig. 4(a)). However, if the driving current density is larger than $j_c$, the AFM skyrmion will arrive at the right terminal of the longitudinal nanotrack, as shown in Fig. 4(b). Thus, the logic operation of AND gate "$1 \cdot 0 = 0$" and the logic operation of OR gate "$1 + 0 = 1$" can be implemented by manipulating the motion of AFM skyrmions in the proposed model. Similarly, if the AFM skyrmion is initially created at the input B side, it will move downward to the lower terminal when $j < j_c$ or move to the output side for $j > j_c$. So, the logic operation of AND gate "0 · 1 = 0" and the logic operation of OR gate "0 + 1 = 1" can also be implemented.

When two AFM skyrmions are separately created in the two input terminals A and B, by controlling the driving current density ($j < j_c$) and the VCMA gradient, the AFM skyrmion from the input A side and the AFM skyrmion from the input B side could arrive at the junction area at the same time. Due to the repulsion between two skyrmions, the right one will be squeezed out of the voltage-gated region by the left one. As a result, the right AFM skyrmion can reach the output side driven by the spin current, while the left one will move to the lower terminal. When $j > j_c$, the left skyrmion could annihilate at the junction area due to the geometric confinement (i.e., the skyrmion-edge interaction). Therefore, the logic operation of AND gate "1 · 1 = 1" and the logic operation of OR gate "1 + 1 = 1" can be implemented. As shown in Fig. 4 (see Supplementary Video 1 and Video 2), the trajectories of the two AFM skyrmions are indicated by yellow lines. The corresponding truth tables of the logic AND and OR gates are given in Fig. 4(c-d). If there are no skyrmions at both two input sides, the output of "0" will be obtained. It should be noted that the delay $\Delta t$ in applying the current is used in our simulation to make two skyrmions of the two branches appear at the intersection of this nanotrack simultaneously. Indeed, the time interval is dependent on the driving current density, external electric field strength and the size of the nanotrack. Therefore, $\Delta t$ could be adjusted by changing above parameters and even reach 0. On the other hand, a voltage bar or voltage gates (which can produce a potential barrier introduced by perpendicular anisotropy) can also be applied in one branch to synchronize skyrmions [39].

Here, we provide the phase diagram of the logical operation as a function of the PMA gradient and the driving current density [see Fig. 5(a)]. The critical driving current density required to convert the logic AND gate to OR gate is also shown in Fig. 5(b). Therefore, one can adjust the driving current density to realize the function of different logic gates in terms of the transition value for different PMA gradients. For each kind of logic gate, the driving current density is constant. When the logic operation of AND gate is realized, the driving current density in the longitudinal nanotrack must be smaller than the critical current density $j_c$, while the OR gate requires the current density greater than $j_c$.

Based on the above design concept, the logic operation of NOT, NAND and NOR gates can also be implemented, which are demonstrated in supplementary materials. For the read-out of information, one can take advantage of the AFM MTJ sensor at the output side. A large tunneling anisotropic magnetoresistance (TAMR) effect has been found experimentally in AFM tunneling junction [63], which makes it possible to indirectly detect AFM skyrmions through MTJ. In addition, when each logic operation is completed, a large current pulse can be applied to erase the AFM skyrmions in the nanotrack. Alternatively, we can also add a vertical track crossing at the output region to transfer the output skyrmions to the external area.

We now discuss the advantages of the proposed design. Firstly, the magnetic skyrmion is a topologically protected spin texture, and due to the fact that the AFM materials are not sensitive to external magnetic fields, the AFM skyrmion-based logic computing gates are promising for future non-volatile memory and have a strong protection against external magnetic disturbances. Also, the AFM skyrmions as information carriers in this model are jointly controlled by the spin-polarized current as well as the VCMA gradient, which can move along the driving force without showing the detrimental skyrmion Hall effect [64]. Therefore, the proposed scheme can be manipulated artificially in a larger working window, which is different from the reported designs [31,36-39] that strongly rely on the skyrmion Hall effect determined by the driving current density and the racetrack geometries. Compared with the already existed logic device based on AFM skyrmions [22], our proposal can realize the all-skyrmion logic operation and avoid the conversion between the domain wall and the skyrmion, thus reducing the energy consumption and simplify the structural design of the racetrack. Furthermore, the energy costs for logic AND and OR gates are $5.832\times10^{-3}$ fJ and $1.382\times10^{-2}$ fJ, respectively, which are two or three orders of magnitude smaller than that required for other comparable skyrmionic logic devices [37,38,65].

In conclusion, we have investigated the dynamic behaviors of a single AFM skyrmion and two AFM skyrmions on a straight nanotrack with a voltage-gated region. We also discussed the dynamics of AFM skyrmions manipulated by electric currents and fields simultaneously in a cross-shaped nanotrack, where the dynamic behavior of skyrmions can simulate well the logic operations of five basic logic gates, including AND, OR, NOT, NAND and NOR gates. Our results are useful for the understanding of AFM skyrmions physics in confined geometries, and can provide useful guidelines for the design and development of future AFM skyrmion-based non-volatile logic computing devices.

**SUPPLEMENTARY MATERIAL**

See supplementary material for the numerical simulation details, the relationship between the transition speed and the voltage applied in the simulation, the demonstration for the logic NOT, NAND and NOR gates based on AFM skyrmions and the energy consumption analysis. Supplementary material videos show the basic Boolean logic operations of AND, OR, NAND, NOR gates realized by manipulating the skyrmion motion in AFM nanotracks.

**AUTHOR'S CONTRIBUTIONS**

All authors contributed intellectual input of ideas and assistance to this study. X. Liang. and J. Xia contributed equally.

**ACKNOWLEDGEMENTS**

G.Z. acknowledges the support by the National Natural Science Foundation of China (Grant Nos. 52111530143 and 51771127) of China, the Scientific Research Fund of Sichuan Provincial Education Department (Grant Nos. 18TD0010 and 16CZ0006). X.Z. acknowledges the support by the National Natural Science Foundation of China (Grant No. 12004320), and the Guangdong Basic and Applied Basic Research Foundation (Grant No. 2019A1515110713). M.E. acknowledges the support by the Grants-in-Aid for Scientific Research from JSPS KAKENHI (Grant Nos. JP18H03676, JP17K05490) and the support by CREST, JST (Grant Nos. JPMJCR16F1 and JPMJCR20T2). O.A.T. acknowledges the support by the Australian Research Council (Grant No. DP200101027), NCMAS grant, and the Cooperative Research Project Program at the Research Institute of Electrical Communication, Tohoku University. X.L. acknowledges the support by the Grants-in-Aid for Scientific Research from JSPS KAKENHI (Grant Nos. JP20F20363, JP21H01364, and JP21K18872). Y.Z. acknowledges the support by Guangdong Special Support Project (2019BT02X030), Shenzhen Peacock Group Plan (KQTD20180413181702403), Pearl River Recruitment Program of Talents (2017GC010293) and National Natural Science Foundation of China (11974298, 61961136006).

**DATA AVAILABILITY**

The data that support the findings of this study are available from the corresponding authors upon reasonable request.

**FIGURES**

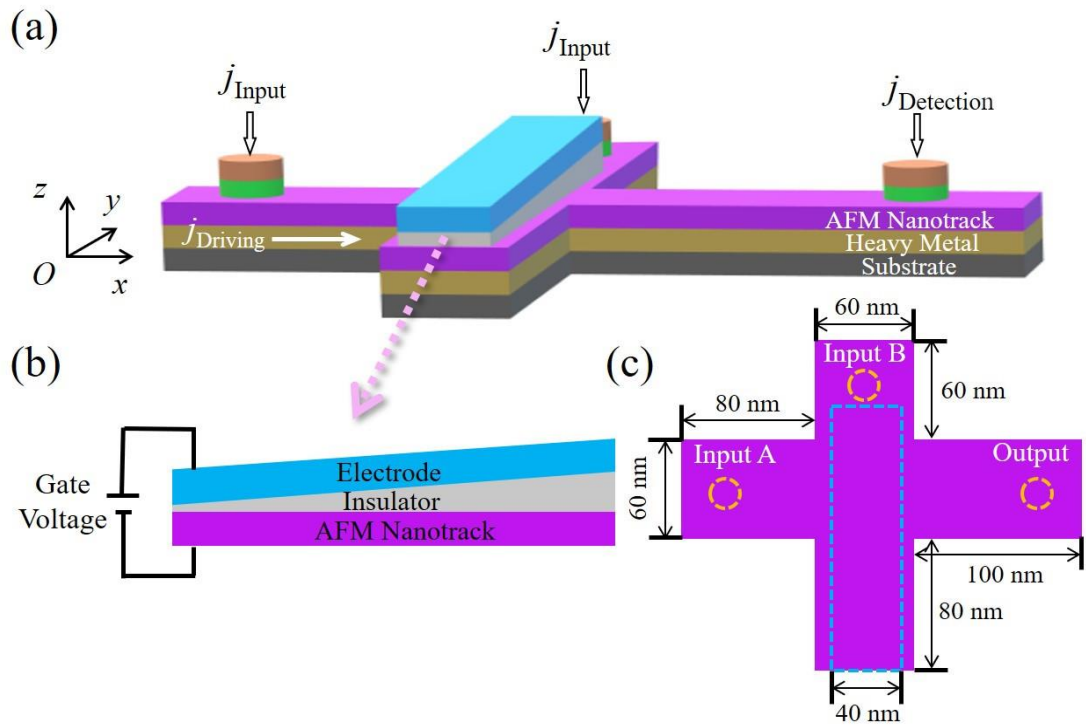

Fig. 1. (a) The 3D illustration of the AFM skyrmion logic AND/OR gates. (b) Schematic view (*yz*-plane) of the transverse nanotrack design. (c) Schematic view (*xy*-plane) of the logic gates, where the yellow dashed circles represent the input and output sensors, and the region indicated by the blue dashed box has a PMA gradient along –*y*-direction induced by the VCMA effect.

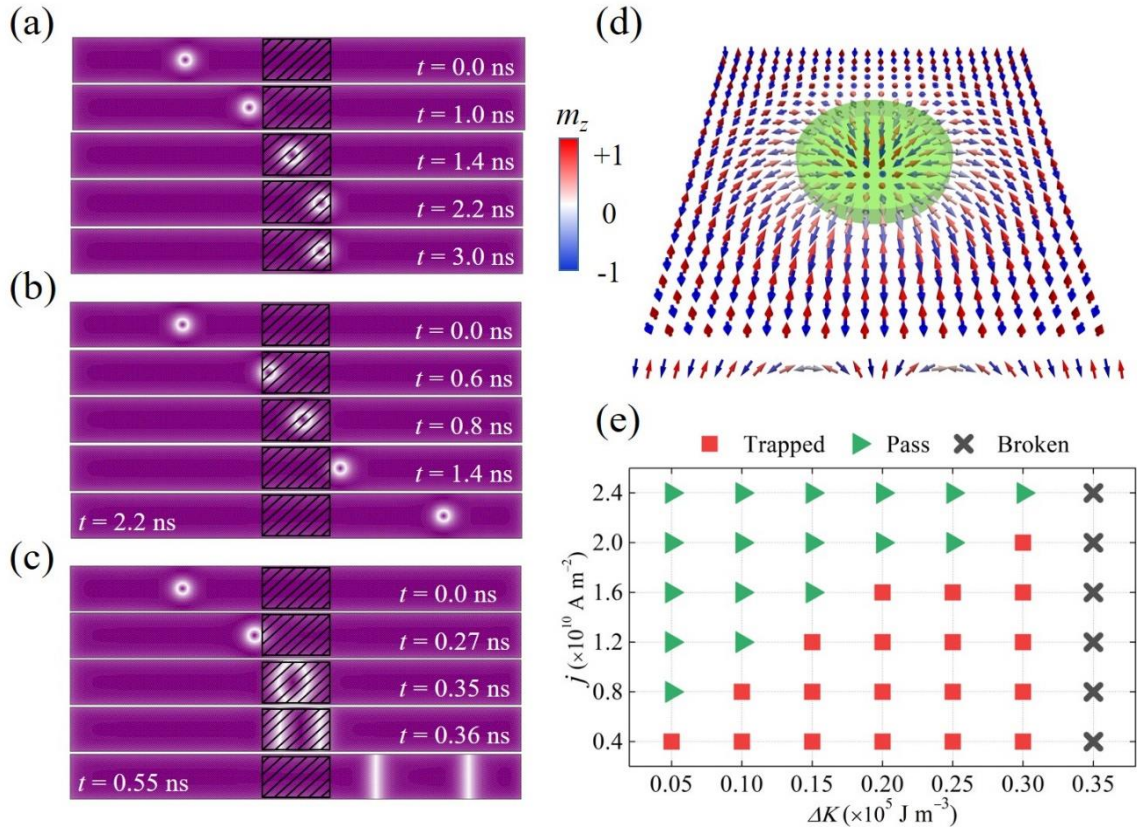

Fig. 2. Current-induced motion of a single skyrmion in an AFM nanotrack with a 60-nm-long voltage-gated region. (a) $j = 1.0\times10^{10}$ A m$^{-2}$, $\Delta K = 0.2\times10^5$ J m$^{-3}$. (b) $j = 2.0\times10^{10}$ A m$^{-2}$, $\Delta K = 0.2\times10^5$ J m$^{-3}$. (c) $j = 4.0\times10^{10}$ A m$^{-2}$, $\Delta K = 0.35\times10^5$ J m$^{-3}$. (d) Illustration of an AFM skyrmion spin texture with the two antiparallel magnetic sublattices. The green cylinder denotes the core area of skyrmion. (e) The working window of the current-induced motion of a single AFM skyrmion. In the simulation, the AFM racetrack with size of $400 \times 40 \times 0.4$ nm$^3$ is adopted.

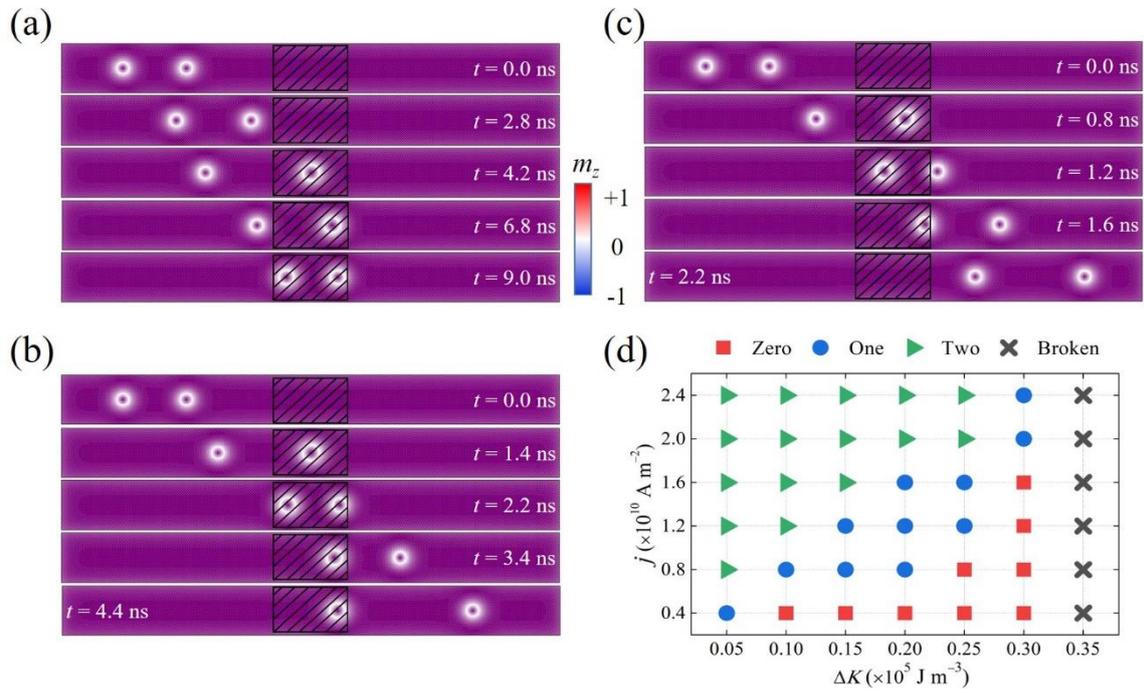

Fig. 3. Current-induced motion of two skyrmions in an AFM nanotrack with a 60-nm-long voltage-gated region. (a) $j = 0.3 \times 10^{10}$ A m$^{-2}$. (b) $j = 1.0 \times 10^{10}$ A m$^{-2}$. (c) $j = 2.0 \times 10^{10}$ A m$^{-2}$. (d) The working window of the current-induced motion of two AFM skyrmions. In (a−c), the difference between $K_v$ and $K_0$ is fixed at $0.2 \times 10^5$ J m$^{-3}$.

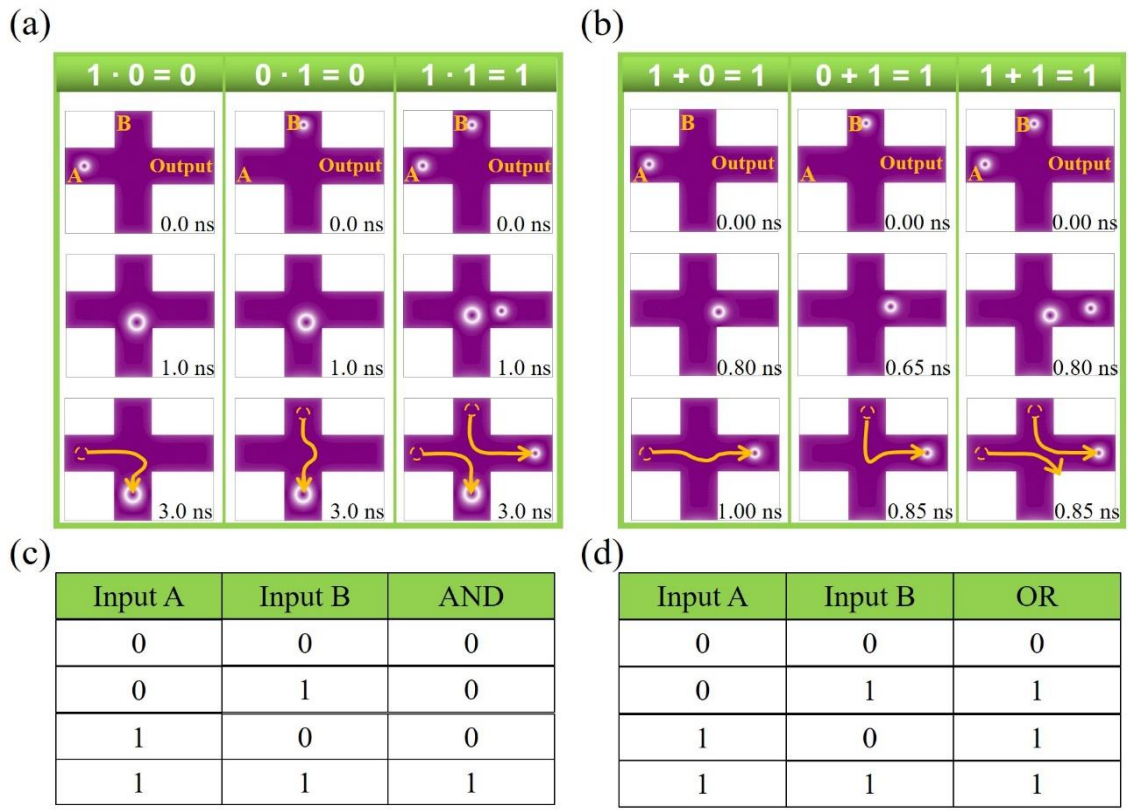

Fig. 4. (a−b) The micromagnetic simulations of the three basic logical operations for the logic AND and OR gates, where the trajectories of skyrmions in AFM nanotrack are identified by yellow lines and the dashed circles represent the initial positions of AFM skyrmions. (c−d) The truth tables of the logic AND and OR gates. In the simulation, $j = 1.5 \times 10^{10}$ A m$^{-2}$, $dK/dy = 400$ GJ m$^{-4}$ and the delay in applying the current is $\Delta t = 0.4$ ns for AND gate. $j = 4.0 \times 10^{10}$ A m$^{-2}$, $dK/dy = 400$ GJ m$^{-4}$ and $\Delta t = 0.48$ ns for OR gate.

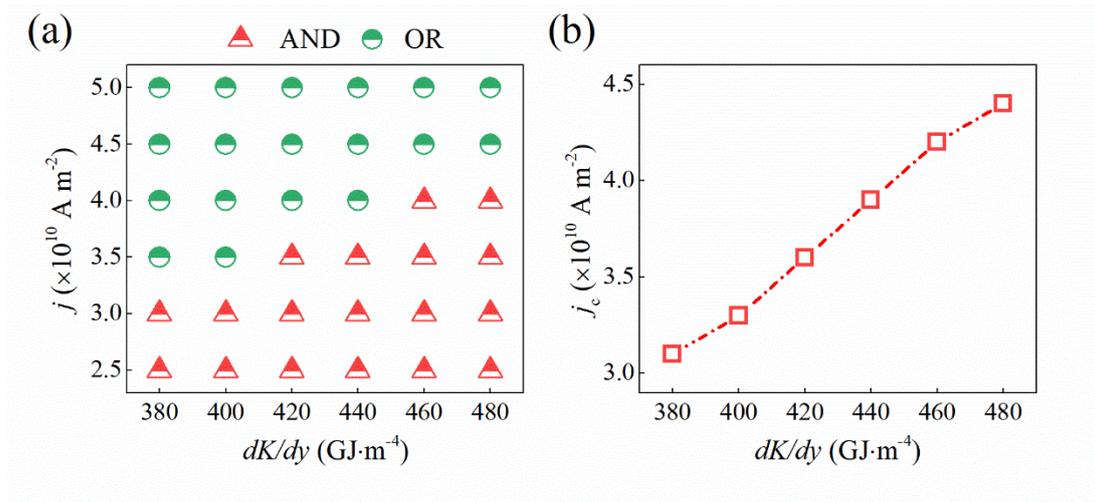

Fig. 5. (a) Logic AND/OR gates as a function of the driving current density and the VCMA gradient. (b)The critical driving current density by which the logic gate changes from AND gate to OR gate.